\documentclass[12pt]{iopart}

\usepackage{graphicx}
\usepackage{amssymb}
\usepackage{tikz}
\usepackage{hyperref}
\usepackage{epstopdf}
\usepackage{verbatim}
\newcommand{\blacke}{\tikz\draw[black,fill=white] (0,0) circle (.5ex);}
\newcommand{\blackf}{\tikz\draw[black,fill=black] (0,0) circle (.5ex);}

\begin{document}

\title{Logical Gates Embedding in Artificial Spin Ice}
\author{Francesco Caravelli and Cristiano Nisoli}
\address{Theoretical Division and Center for Nonlinear Studies,\\
Los Alamos National Laboratory, Los Alamos, New Mexico 87545, USA}
\ead{caravelli@lanl.gov}
\vspace{10pt}

\begin{abstract}
The realization and study of arrays of interacting magnetic nanoislands, such as artificial spin ices, have reached mature levels of control that allow design and demonstration of exotic, collective behaviors not seen in natural materials. Advances in the direct manipulation of their local, binary moments also suggest a use as nanopatterned, interacting memory media, for computation {\it within} a magnetic memory. Recent experimental work has demonstrated the possibility of building logic gates from clusters of interacting magnetic domains, and yet the possibility of large scale integration of such gates can prove problematic even at the theoretical level. Here we introduce theoretically complete sets of logical gates, in principle realizable in an experiment, and we study the feasibility of their integration into tree-like circuits. By  evaluating the  fidelity control parameter between their collective behavior and their expected logic functionality we determine conditions for integration. Also, we test our numerical results against the presence of disorder in the couplings, showing that the design gate structure is robust to small coupling perturbations, and thus possibly to small imperfections in the fabrication of the islands.
\end{abstract}

\section{Introduction}
The last fourteen years have seen the use of interacting~\cite{colloq,Wang1} magnetic nanostructures~\cite{Bader} patterned in different geometries in so called artificial spin ices, to realize a wealth of different emergent behaviors often not found in natural magnets. The level of control afforded by these materials has allowed the study of frustration and residual entropy, within a broad range of exotic phenomena and potential applications~\cite{ reddim,Heyderman,Canals1,reddim,Nisoli1,Cugliandolo2,Morgan,Budrikis,Branford,Ryzhkin,Castelnovo1,Ladak1,topor,Chern2,Le,Chern3,Gliga,Nisoli4,Gilbert2,Bhat}.
Recently, Gartside et al. and Wang et al. have demonstrated the fine, local manipulation of islands magnetization \cite{gartside,WangYL2,WangYL} to reliably write the bites of artificial spin ice~\cite{colloq,Nisoli4} materials. Moreover, still unpublished results~\cite{vavassori0,vavassori}  demonstrate local activation of the islands kinetics via photo-induced heating. The use of artificial spin ices as {\it collective} rather than individual forms of memory has also been explored~\cite{GilbertMem,Lammert2}.
 
 This confluence of results opens now intriguing perspectives toward new platforms for computation within a memory medium. In current magnetic memory storage, it is important that each magnetic domain is not influenced by the other domains to prevent data corruption. However, an interacting memory could allow for computation performed {\it within} the memory~\cite{chialvo,diventra,traversa}. In a network of interacting memory bits there would be no obvious directionality of flow from input to output within the logic gates composing it. It has been proposed that, in the absence of a logic unit, information overhead (the capability of compressing information in the collective state of the network of magnets), intrinsic parallelism, and functional polymorphism 
would be natural \cite{memr1}. 

 In a network of interacting, binary spins, some could be kept fixed in time and considered inputs via external fields or via nearby larger islands used as biases. Other islands could be considered outputs or results intermediate computations, to be read after the system has relaxed in a low-energy state; this is some other fixed point, which would correspond to the result of a computation. Indeed, Boolean gates of interacting, magnetic nanoislands have been already pioneered~ \cite{bernstein1,bernstein2,bernstein3} in the context of nonstandard logic and more recently within artificial spin ice materials~\cite{Arava, CASI}. However, the crucial issue neglected in most experimental work is the reliability of their integration in realistic circuits, which can be problematic~\cite{bernstein4,Gypens,Arava}. 

In this work 
we  explore theoretically  the feasibility of integrating nanomagnetic gates in functional circuits  by simulating thermal annealing of  tree-like circuits made of magnetic nanoislands. Previous work in this respect has been done via the implementation of boolean gates directly in the couplings \cite{Biamonte}. However, in this work we consider the possibility of reprogramming the gates \textit{a posteriori} using  intermediate islands exerting a biasing field  (see below). This prompts us to perform a new analysis in order to go beyond previous works on the Husimi cactus \cite{Jurcisinova}. We will also consider localized, rather than homogeneous, annealing that proceeds along the circuit, now a possibility via photo-activation~\cite{vavassori0}. The purpose of this paper is to verify the {\it fidelity control parameter} between the convergence of such tree-like interacting systems and the logical functionality of the corresponding logic circuits. 

\section{Gates}
It is not difficult to devise logical gates from interacting nanoislands whose magnetization can be modeled as a classical, binary Ising spin. 
For  definiteness, consider the NAND and NOR gates (which are universal gates) realized via triangular plaquettes of antiferromagnetically interacting classical Ising spins in Fig.~1 which can be (and have been, though in a different context~\cite{Ke}) fabricated at the nanoscale with magnetic moments perpendicular to the array. We imagine that red spins can be set as inputs and held fixed during computation, while the green spin is the output to be read. The blue spins provide a bias, that is a local field $h$ on the output that eliminates any indeterminacy of the output  that results from frustration. Its orientation determines whether the gate is (N)AND or (N)OR, and allows for reconfigurable circuitry. 
In practice these gates can be realized through a nanoisland of higher coercivity, such that its moment does not change during computation. Switching the magnetization of such an island leads to changing between (N)AND and (N)OR, thus reprogramming the circuit.

One can then integrate gates into circuits. As an example, Fig. \ref{fig:fibonacci} shows integration to produce a 3-bits circuit meant to select Fibonacci numbers. 

\begin{figure}
\centering
\includegraphics[width=.6\columnwidth]{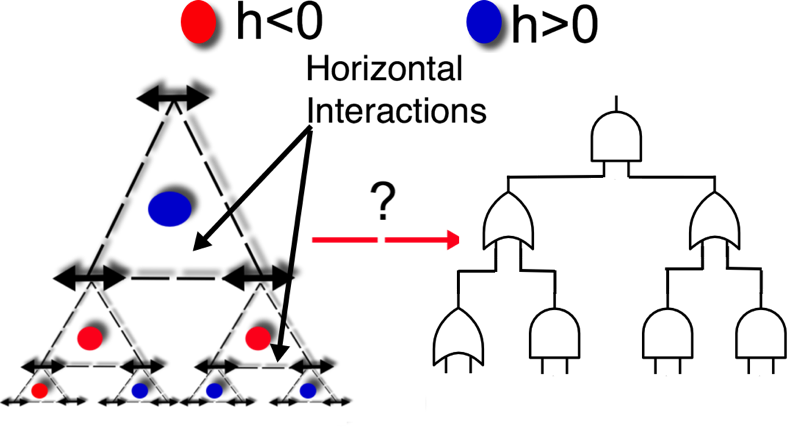}
\caption{ Schematics of a  circuit associated with a local magnetization on the triangles (dashed lines represent the interactions, double-arrows the fluctuating spins, blue/red dots the orientation of bias). We highlighted what we called horizontal interactions.}
\label{fig:gates}
\end{figure}

\begin{figure}[!t]
\center
\includegraphics[width=.5\columnwidth]{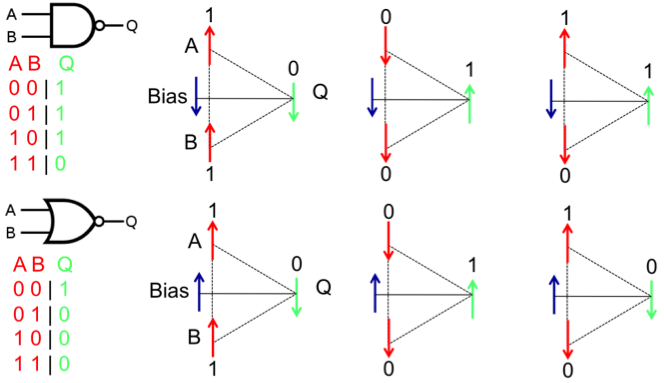}
\caption{A magnetic NAND gate (left) and NOR gate (right) built from antiferromagnetically interacting moments, and their truth table in Boolean algebra and their realization through antiferromagnetic spins. In red we have the input spins, in green the output. The blue spin is needed to bias frustration and changes the plaquette from a NAND  to a NOR, making the gate re-programmable.}
\label{fig:fibonacci}
\end{figure}

\begin{figure}
\centering
 \includegraphics[width=.6\columnwidth]{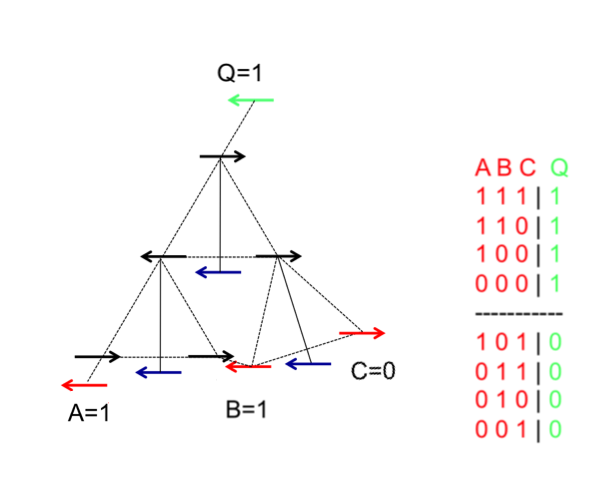}
\caption{ A spin-circuit to solve the Fibonacci sequence.}
\label{fig:fibonacci2}
\end{figure}

The goal of this paper is to implement a logic gate structure, such as the one of  Fig. \ref{fig:gates}, in spin ice systems. In Fig. \ref{fig:fibonacci}, we have shown as an example the case of antiferromagnetic bits; in reality, much more freedom is allowed in choosing the couplings among spins, and therefore in engineering gates, as we show below. Spins can be perpendicular to the array or in plane, and in both cases, have been realized in practice~\cite{Ke}.  When spins are in plane, because of the anisotropy of the dipolar interaction, the different relative orientation of the magnetic island can be translated in different coupling constant, ferro- or antiferromagnetic. Indeed it is even possible to remove the coupling between the inputs of a gate setting the moments perpendicularly. As an example, in Fig.\ref{fig:fibonacci2} we represent a Fibonacci series solver using such mapping.
In fact, any (N)AND, (N)OR gates can be realized in practice (see Supp. Mat. for a more general analysis).

\begin{figure}[t!]
\center
\includegraphics[width=.7\columnwidth]{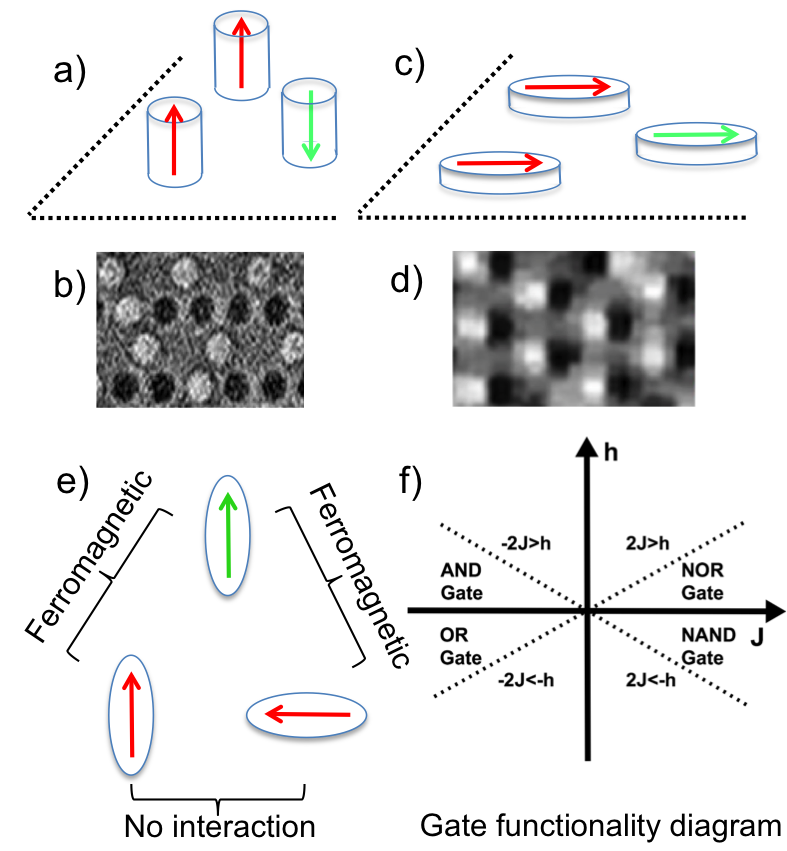}
\caption{Gates of different properties can be fabricated in different ways. (a) Schematics of out of plane nanoislands leading to magnetic moments  antiferromagnetically coupled. (b) MFM images of out of plane magnetic moments made of lithographically fabricated nanoislands~\cite{Zhang2} (c) Schematics of in plane nanoislands leading to ferromagnetic coupling. (d) MFM images of in plane magnetic moments made of lithographically fabricated nanoislands~\cite{Ke}. (e) Example of suppression of the coupling between input bites, using a $90\deg$ arrangement for input nanoisland, in plane. (f) Schematics of gates that can be obtained in the $(J,h)$ parameter space in the ground state of the spin Hamiltonian of eqn. (\ref{eq:paramH2}).}
\label{fig:paramH}
\end{figure}

The gate functionality for three spins can be obtained by studying a simple spin Hamiltonian of the form
\begin{equation}
H_3=\sigma_3\left( J(\sigma_1+\sigma_2)+h\right),
\label{eq:paramH2}
\end{equation}
in which we assume that the spins $\sigma_1$ and $\sigma_2$ are inputs, and $\sigma_3$ is our output and free to fluctuate at a certain temperature $T$. However, it is not hard to see that if $2|J|>h$ the ground state of the system are those of a logic gates (N)AND and (N)OR, where the negation N depends on the ferro- or antiferromagnetic coupling, and the type of gates on the biasing field $h$.
In Fig. \ref{fig:paramH} (f), we show the gate functionality as a function of the parameters $J$ and $h$, keeping in mind that we assume the underlying structure to be based on magnetic nanoislands and artificial spin ices.

\section{Integration}

While gates are easy to conceive in a system of interacting binary variables, there are, however, various problematic aspects in going from a single gate to a circuits. One is the tolerance to errors in the input states, which we discuss in Supp. Mat. A more relevant problem, however, is degeneracy induced by frustration and long-range interactions and/or disorder. Note that gates were introduced under the assumption that the input is held fixed while the output is computed. That cannot be done for gates inside a circuit, where one gate's input is another gate's output. Also, in the absence of frustration or long-range interactions,  if the strength of couplings is the same along the hierarchy,  then some spins at intermediate stage can arrange itself \textit{without} changing the output (see Sup. Mat.). While this might seem irrelevant, logical fidelity not only of output, but of each constitutive gate is necessary to go beyond a Turing approach toward neuromorphic computing in a network circuit. 

This problem can be solved by modulating  the couplings $J$ among islands and make them  scale with the layer of the tree, which we label by $k$. One possibility is to choose $J_{k+1}={J_k}/({2+\epsilon})$ for some $\epsilon>0$ and $h_{k+1}={h_{k}}/{2}$ where $h$ is the local field from the biasing island which selects gate functionality. For the model without horizontal (input) spin interactions, we choose $\epsilon$ such that $|{J_k}/({2+\epsilon})|<|h_{k }|$ and 
\begin{equation}
|J_{k-1}|\leq \frac{|J_{k}|}{(2+\epsilon)}.
\label{eq:coppresc}
\end{equation}

We perform numerical simulations of randomly chosen gates via a Glauber~\cite{redner} spin dynamics with exponential annealing. 
The probability of a spin flip leading to an energy change $\Delta E$ is given by
$p(k)=(1+e^{\Delta E /T_k(t)})^{-1}$, 
where $T_k$ is the temperature at the layer $k$. We perform both uniform annealing, corresponding to $T_k(t)=T_0 \lambda^t$, with  $\lambda<1$ a measure of how slow the annealing is (slower for $\lambda$ closer to 1), but also a layer dependent temperature to simulate light heating~\cite{vavassori0,vavassori} or $\tilde T_k(t)=T_0 \lambda^t/2^k$: the lower layer (closer to inputs) will cool first, and the top layer last.

\begin{figure}[t!]
\center
\includegraphics[scale=0.38]{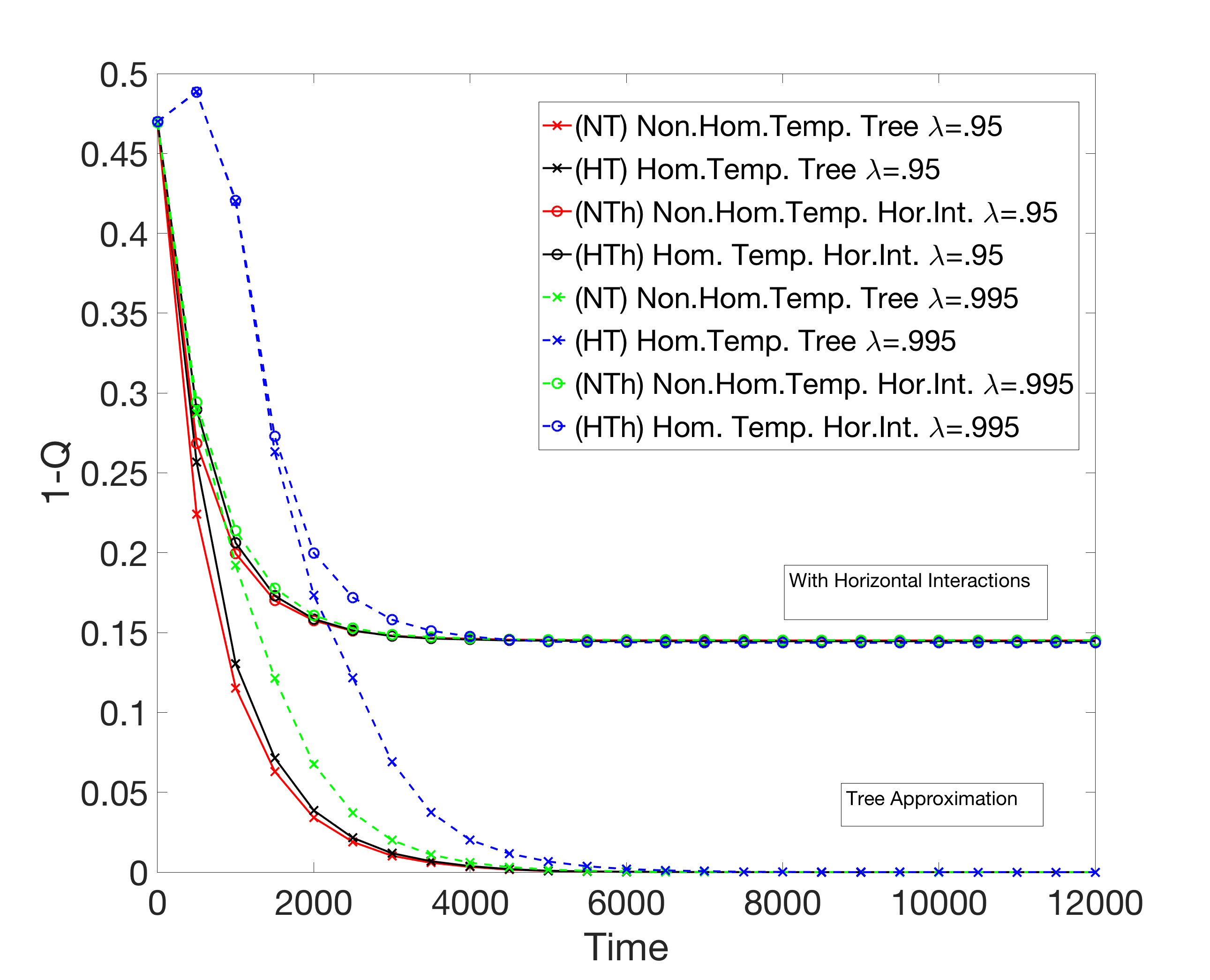}
\caption{ Behavior of the  fidelity control parameter as a function of time when the system is annealed. We denote with mark $x$ curves belonging to exact trees and $o$ curves with horizontal interactions.
The curves are obtained after averaging over 100 samples of random gates with $n=10$ layers (2048 thermally activated spins and 2048 inputs) random gates. We observe that while for a tree the curves relax to $\mathcal Q=1$, and the trees with horizontal interactions relax to an approximate value of $\mathcal Q\approx 0.85$, thus presenting a macroscopic number of defects. 
Here $1$ time unit is $N$ Monte Carlo flip attempts, with $N$ the number of spins. We also observe that in the case with homogeneous temperature, the  fidelity control parameter relaxes a bit slower compared to the non-homogeneous case. In the legend, $\lambda$ is the annealing rate, e.g. $T_k=\lambda T_{k-1}$, and thus for $\lambda=1$ no annealing occurs.} Legend:  homogeneous temperature on a tree (HT), non-homogeneous temperature on a tree (NT), non-trees with homogeneous (HTh) and non-homogeneous temperature (NTh). In the legend, $N$ stands for nonhomogeneous while $H$ for homogeneous, and $T$ for temperature; $h$ stands for horizontal interactions added. A video of the annealing is provided in \cite{youtube}. 
\label{fig:annealing}
\end{figure}
As a control parameter for the fidelity of the circuit we introduce   a control parameter to measure the difference between the spin system and the equivalent logic circuit, not only for final output, but  {\it for all the intermediate layers}. For each gate, we consider the vector of spin orientations $\vec L$ corresponding to expected logical functionality, and the one obtained from the interacting spin system, $\vec S$. Using the vector $\vec L$, we can define the \textit{gate fidelity control parameter} as the quantity
\begin{equation}
\mathcal Q=\frac{\left[\vec S\cdot\vec L+(1-\vec S)\cdot(1-\vec L)\right]}{N},
\end{equation} 
where $N$ is the total number of gates. Full functionality corresponds to $\mathcal Q=1$, and a system in which all gates fail has $\mathcal Q=0$. Completely random (high temperature) systems have $\mathcal Q=0.5$, as an uncorrelated gate has probability 0.5 of being in the right output state. We call this overlap {\em total}. Later we will also use an {\em output} overlap, where only the outputs are considered and not the intermediate spins. 

We consider two kinds of trees. One kind lacks  {\it horizontal} interactions (i.e. those between the inputs of gates) annealed homogeneously (HT) and non-homogeneously (NT). The other has horizontal interaction, and also is annealed  homogeneously (HTh) and non-homogeneously (NTh). Trees have $n=12$ layers of gates, corresponding to 2048 total gates, 2048 floating spins, and input of 2048 bits (e.g. 4096 total spins between inputs and thermally activated spins). On each of the four case we perform 100 simulations corresponding to 100 different circuits (that is, assigning randomly reprogramming biases in each gate) with 100 different random inputs. 

Fig.~\ref{fig:annealing}  shows the average gate overlap for the four cases. In absence of horizontal interactions,  the system converges to the proper logical output with {\it total} overlap one. This {\em demonstrates gates integration for deterministic computing}.

\begin{figure}[t!]
\center
\includegraphics[width=.7\columnwidth]{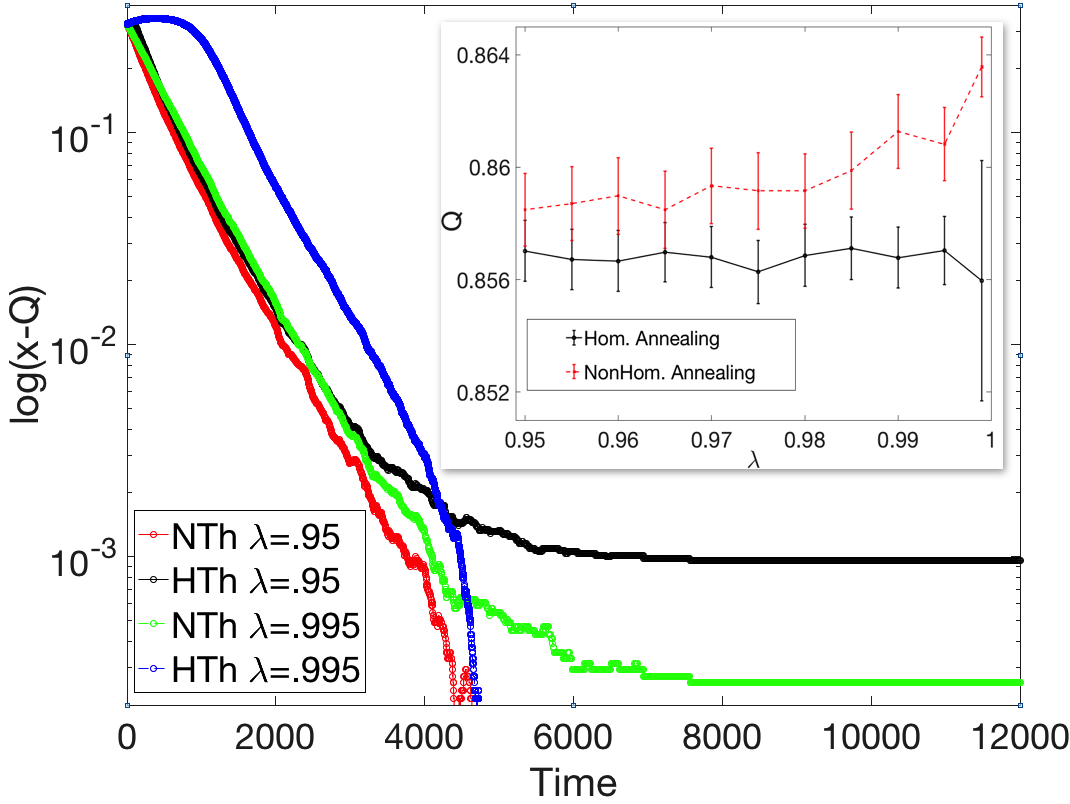}
\includegraphics[width=.7\columnwidth]{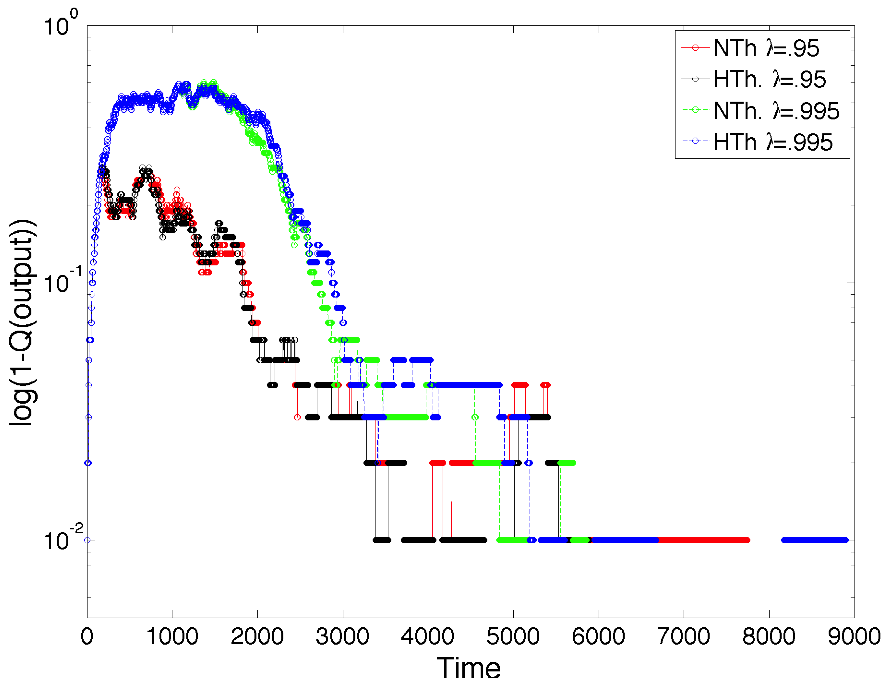}
\caption{Plots of the convergence of the overlap parameter for homogeneous (HT) versus non-homogeneous temperatures (NT) for the case with horizontal interactions and for two annealing rates. Top: Log-plot of the $x-\mathcal Q$, showing an approximate exponential relaxation ($x$ is the best value obtained). Inset: convergence value of ${\mathcal Q}$ for various annealing rates $\lambda$ shows approximately constant faultiness in the homogeneous annealing, but also their systematic reduction in the nonhomogeneous annealing. Bottom: Gate overlap evaluated numerically only on the output result, which represents the fidelity of the final computation. We see that despite these having a non-zero probability of not going into the ground state, this is remarkably low even for the case with horizontal interactions for a number of layers $n=10$. Here $1$ time unit is $N$ Monte Carlo flip attempts, with $N$ being the number of spins.}
\label{fig:annealing2}
\end{figure}

Instead, in the case of horizontal interactions, ${\cal Q}$ does not converge to  one, and the system performs at only about 85\% accuracy.  Furthermore, as Fig.~\ref{fig:annealing2} (top) shows, a noticeable difference  can be observed between the curve of homogeneous and non-homogeneous annealing. The non-homogeneous case converges considerably faster. Also,  Fig.~\ref{fig:annealing2} (bottom) shows that a non-homogeneously annealed tree with horizontal interaction converges to slightly higher values of overlap for slower annealing, unlike the homogeneous annealing. 

It is important to note that the measure $\mathcal Q$ is a quite restrictive one. One might argue that the result of the computation might still be correct when the intermediate computations contains faulty gates.  One can in fact introduce another measure of overlap $\mathcal Q(output)=s_{out}l_{out}+(1-s_{out})(1-l_{out})$, where $l_{out}$ is the expected final result of the computation and $s_{out}$ is the fluctuating (output) spin in our Monte Carlo simulations. It is interesting to note that we observe numerically $\langle \mathcal Q(output)\rangle >\langle \mathcal Q\rangle$. Figure~\ref{fig:annealing2}, bottom, shows that  while loop-containing tree circuits overlap with predicted logic functionality at only $\sim 85\%$, the overlap of the output is still very close to 1, meaning that that internal defects seem to have a finite correlation length (e.g. do not propagate to the output).
While loops are detrimental for the general efficiency of the embedding, they do not seem to affect the output.

\begin{figure}
    \center
    \includegraphics[scale=0.3]{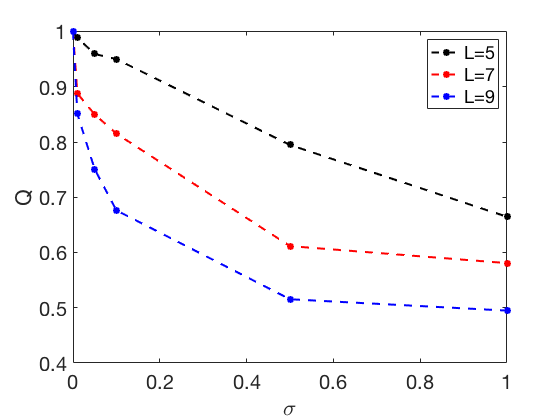}
    \caption{Fidelity control parameter as a function of the noise variance strength $\sigma$ for $n=5,7,9$. As we see, in both cases we obtain that the ovelap is reduced, and already for $n=9$ and $\sigma=1$ the  fidelity control parameter reaches the completely random value $\mathcal Q=0.5$.}
    \label{fig:noise}
\end{figure}

Finally, we tested robustness of computation against quenched disorder. This can come from fabrication, but also from the next nearest neighbor interactions. The strength and signs of these interactions depend on the relative orientation of the islands, and thus not only on the topology but the overall orientation of the circuit. In order to test the role of spurious interactions, we can introduce noise in the couplings and see how this effects the degree of computation. For this purpose, to the overall coupling matrix we add a Gaussian independent and identically distributed noise $J^{\sigma}_{ij}=\mathcal N(0,\sigma)$, e.g. random noise with zero mean and variance $\sigma$. In Fig. \ref{fig:noise} we show the  fidelity control parameter as a function of the noise strength $\sigma$, averaged over different realization of the noise.
The results thus show that noise can dramatically affect computation in these logical circuits.

%

We conclude that integration of gates of floating spins into tree-like circuits without loops due to horizontal interactions leads to the theoretical possibility of deterministic computations, whereas loops imply circa 15\%  faultiness. There is, however, fault tolerance in the output.  We interpret this result as an effect of the non-triviality of the energy landscape when the loops are present, and thus to local minima in which the system gets trapped.

\section{Faults, Defects, and Kibble-Zurek}

One of the motivations for computing within the memory lies in overcoming the Turing paradigm. That would entail in general loop-containing networks (which might be used to enforce constraints on computation), and also functional polymorphism~\cite{diventra}, which can blur the difference between inputs and outputs. 
It is therefore relevant to investigate the origin of the internal faultiness seen in interaction graphs containing loops.  

A possible explanation lies in the Kibble-Zurek (KZ) mechanism~\cite{Kibble,Zurek}, which describes the formation of a nonzero density of defects (excitations about the ground state) when a system crosses a critical temperature during an annealing.  The second possibility is a glass transition in the spirit of the random field Ising model, in which the local external fields (our biases) are random. This scenario has been studied in \cite{RFIM} for the case of Cayley trees. There, it was noticed that the random field Ising model has a second order phase transition. The presence of a phase transition is therefore important and worth exploring. As we anneal the system and we cross a transition temperature, defects form at transition either via a coarsening or a Kibble-Zurek mechanism. These defects (and the rate at which they can be reabsorbed) prevent reaching the designed ground state, or in our language, a correct result of the computation.
 
 \begin{figure}[t!]	
 \center
\includegraphics[scale=0.25]{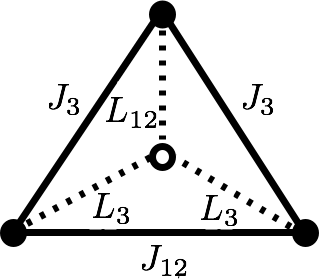}\\
\includegraphics[scale=0.22]{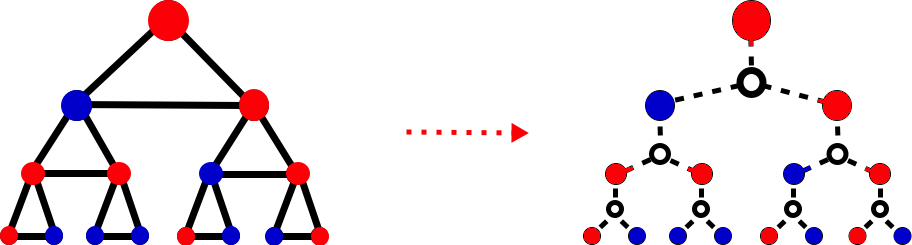}
\caption{The triangle-star map allows to write an equivalent partition function on a tree structure, which can be exactly solved, even in the presence of horizontal interactions.}
\label{fig:isingtreetriangle}
\end{figure}

In order to identify a possible second order phase transition we study the Yang-Lee zeros of the partition function~\cite{YL}, which accumulate toward the real axes of the complex plane (in the thermodynamic limit) when a transition is present. 
Because we are dealing with trees, it is possible to obtain a recursive formula for the partition function by exploiting an old trick, the star triangle relationship, shown in Fig. \ref{fig:isingtreetriangle} (See Appendix). Following for instance Baxter~\cite{Baxter},  we see that each local triangle can be turned into a local tree via 
the introduction of a virtual spin in the gates, with coupling constants $L_{12}, L_3$ given by
\begin{eqnarray}
L_{3}&=&\frac{1}{2} \sinh^{-1}\left(\frac{1}{\sinh(2 J_3) \kappa}\right), \nonumber \\
L_{12}&=&\frac{1}{2} \sinh^{-1}\left(\frac{1}{\sinh(2 J_{12}) \kappa}\right), \nonumber \\
\kappa&=&\frac{(1-v_1^2)(1-v_2^2)(1-v_3^2)}{4\sqrt{(1+v_1 v_2 v_3)(v_1 +v_2 v_3)(v_2 +v_1 v_3)(v_3 +v_2 v_1)}},
\end{eqnarray}
and  $v_1=v_2=\tanh(J_3)$, $v_3=\tanh(J_{12})$.
%
Using these parameters, we obtain a recursion relation for the partition function
\begin{eqnarray}
Z^{n}(\sigma^0,{\tilde h}_{\sigma^0})&=& 2 \cosh({\tilde h}_{\sigma^0}+\tilde L_3) \nonumber \\
&\cdot& Z^{n-1}(\sigma^1,{\tilde h}_{\sigma^1}+\tilde L_{12}) Z^{n-1}(\sigma^2,{\tilde h}_{\sigma^2}+\tilde L_{12}) \nonumber \\
&+&2 \cosh({\tilde h}_{\sigma^0}-\tilde L_3)  \nonumber \\
&\cdot&Z^{n-1}(\sigma^1,{\tilde h}_{\sigma^1}-\tilde L_{12}) Z^{n-1}(\sigma^2,{\tilde h}_{\sigma^2}-\tilde L_{12})\nonumber \\
\label{eq:recz}
\end{eqnarray}
from which zeros can be computed as a function of the length $n$ of the tree (here $\tilde h=h/T, \tilde L =L/T$). 

\begin{figure}[t!]
\center
\includegraphics[scale=0.4]{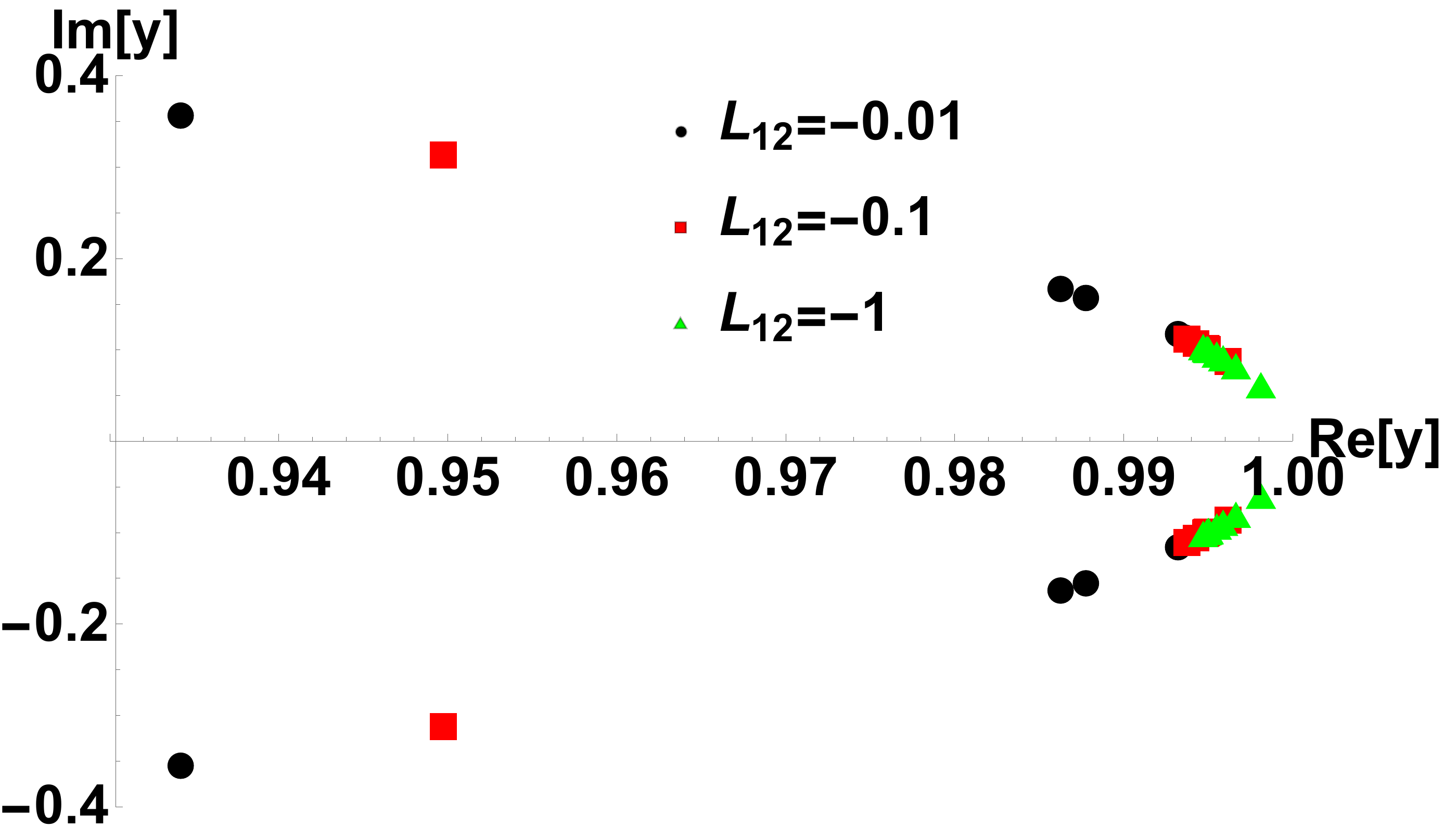}
\caption{Location of the zeros $y=e^{\frac{L_{12}(J}{T}}$ of the partition function in the complex plane, from the recursion relation with horizontal interactions. We calculate the values of the zeros for $J_3=1$, and $J_{12}=0.1$, and we solve numerically for the location of the Yang-Lee zeros, as a function of the number of layers $(n=5\cdots 13)$. The maximum number of layers that we are able to resolve numerically is $n=13$, which can be achieved thanks to the recursion relation. As the number of layers increases, the zeros move to the right and accumulate around $y=1$. This would hint towards the emergence of a phase transition, also for small values of $J_{12}$ and faster for larger values. However, in Fig. \ref{fig:leeyangconv}, we see that the convergence of the zeros stops for a larger number of layers (exponentially many more spins) to a value which is not $1$.}
\label{fig:zeros}
\end{figure}

\begin{figure}
    \center
    \includegraphics[scale=0.37]{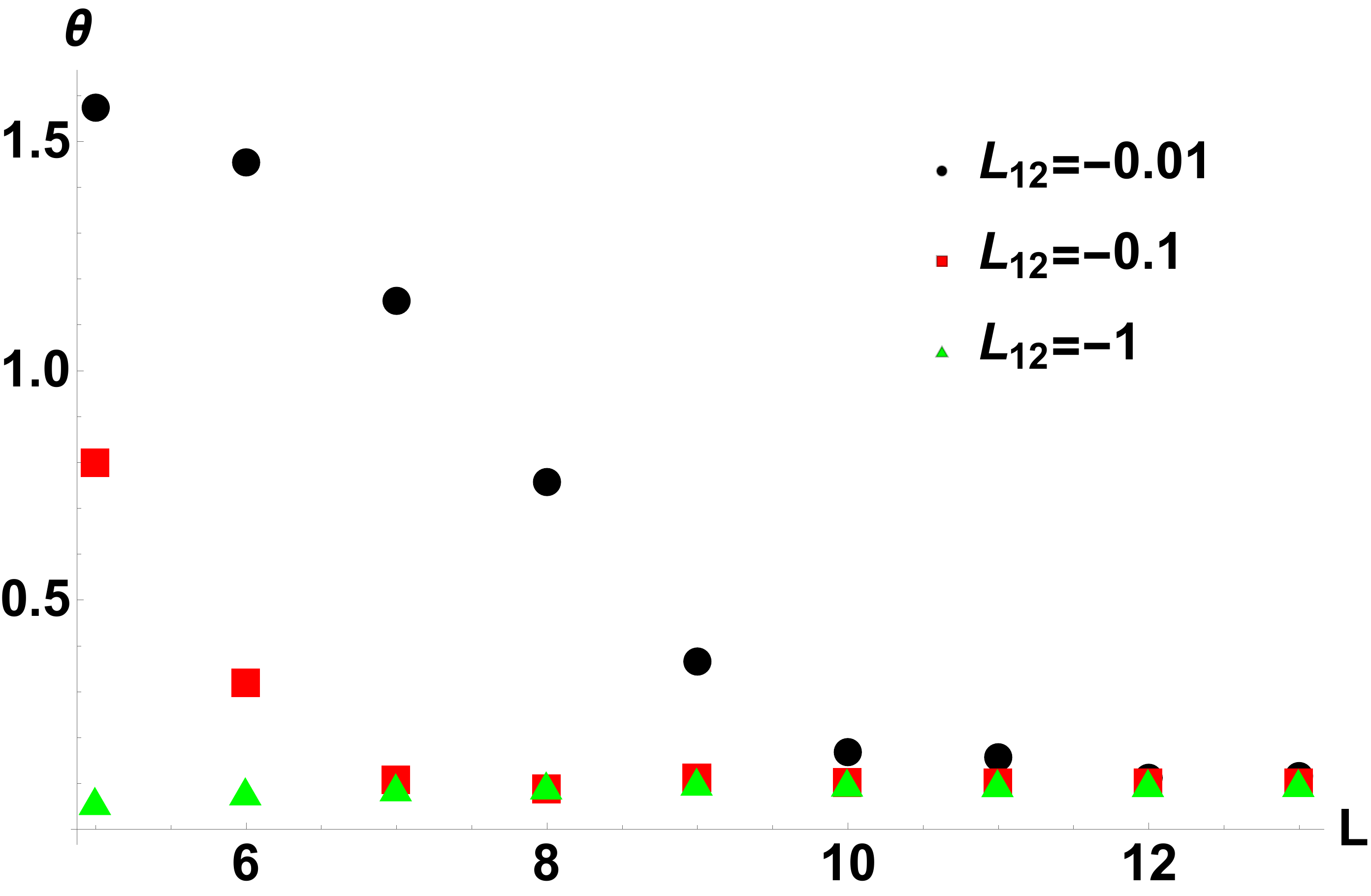}
    \caption{Convergence of the Lee-Yang zeros with the number of layers, and as a function of $T$ for $J_3=1$ and $J_{12}=0.1$. We write $z=e^{i\theta}$, and $\theta\rightarrow 0$ would imply the emergence of a phase transition. We see that as the system increases (exponentially) in size, the zeros do not converge to the zero. In fact, as the system increases, the zeros converge to a finite number in the complex plane.}
    \label{fig:leeyangconv}
\end{figure}

We can thus understand the difference in the annealings of Fig. \ref{fig:annealing} as follows. Whenever  $J_{12} \ne 0$ ,  Fig.~\ref{fig:zeros} shows that the zeros in the complex fugacity plane  $y=e^{\frac{L_{12}}{T}}$ converge to the a value $y^*$ value as $n$ increases for all AND gates and for two values of the temperature, and $J_3=1$ and $J_{12}=0.1$. The value, while being not one, is close to the phase transition value $y^*=1$, as we can see in Fig. \ref{fig:leeyangconv}. We interpret this as a symptom of metastability.
  This is the case for both  horizontal and non-horizontal interactions. However,  with horizontal interactions, the effective size of the system is larger due to the presence of the virtual spins. 
 Thus, in the case with horizontal interaction the system reaches the ground state with more difficulty. For a larger number of layers ($n\geq 9$), the zeros converge to a finite value for various values of the temperature. Another way of interpreting Fig. \ref{fig:leeyangconv} is that in the case of homogeneous trees (all gates identical) one never effectively has a phase transition for a finite system, while it is the case for an infinite one; this is an important difference between the Cayley tree and the Bethe lattice \cite{Baxter}.
 While this is true for ferromagnetic trees (corresponding to AND for instance), it is not true for random gates \cite{RFIM}, in which a second order phase transition is known to exist. 
 Let us use for instance the case of the Kibble-Zurek scaling~\cite{Kibble,Zurek}: the density of defects  scales as $\rho={ \xi}^{-d}$, where $d$ is the dimension of the system and $\xi$ the correlation length at the freezing time, which is related to the linear annealing parameter $\tau$ as $\xi\approx \sqrt{\tau}$. From the exponential annealing, we see that $\tau^{-1}=-T_c \log(\lambda) /J$. For a quasi one-dimensional system, like the one of interest in this paper, we have thus  ${\mathcal Q}=1-\rho\simeq 1-\approx\sqrt{ T_c (1-\lambda)}$ as $\lambda \to 1$ (i.e. infinitely slow annealing). 
 Therefore, in our simulations in which we have random gates, we conjecture that the defects we observe are due to the transition observed in \cite{RFIM}.

\section{Conclusions}
We proposed logical gates made of interacting spin and tested the possibility of integration in tree-like circuits. We  find that in absence of interaction loops and for finite systems, it is possible to mimic deterministic computations by integrating logic gates made of interacting magnetic bits. In the case with loops, the presence of internal, faulty gates, best described in terms of temperature dependent probabilistic gates, suggests directions toward probabilistic~\cite{ProbComp0} rather than deterministic computing. The computation of a specific Boolean configuration can be in general thought of as one particular configuration of random field on a tree, which presents interesting universal behavior~\cite{RFIM} and will be studied elsewhere for our case.
One of the drawbacks of our systems is that it scales exponentially with the number of layers, implying that only a few layers might be practical. Another potential problem in practice is the local, biasing field, necessitating islands of different coercivity. Nonetheless, our analysis cautiously suggests the viability of deterministic computation using interactive, magnetic memory bits, realized by single domain magnetic nanoislands. In future work, we will explore the probabilistic computing aspects of these systems.

\textbf{Acknowledgments.} We would like to thank Fabio Caccioli for comments during the work, and Fabio Lorenzo Traversa for discussion at the beginning of the work. This work was carried out under the auspices of the NNSA of the U.S. DoE at LANL under Contract No. DE-AC52-06NA25396. FC was also financed via DOE-ER grants PRD20170660 and PRD20190195.


\clearpage

\textbf{Bibliography}
\begin{thebibliography}{99}



\bibitem{colloq} C. Nisoli et al.,
Rev. Mod. Phys. 85, 1473 (2013)
\bibitem{Wang1} R. F. Wang et al., 
Nature 439(7074):303-6, (2006). 

\bibitem{Bader} S.D. Bader, 
Rev. Mod. Phys., 78(1):1, (2006). 

\bibitem{reddim} I. Gilbert et al., 
Nature Phys. 12, 162-165 (2016)
\bibitem{Heyderman} L. J. Heyderman, R. L. Stamps, 
J. of Phys.: Cond. Matt., 25(36):363201 (2013) 
\bibitem{Canals1} B. Canals et al., 
Nat. Comm. 7 (2016)
\bibitem{Nisoli1} C. Nisoli et al, 
Phys. Rev. Lett., 98(21):217203 (2007)
\bibitem{Cugliandolo2} D.Levis et al., 
Phys. Rev. Lett., 110(20):207206 (2013)
\bibitem{Morgan} J. P. Morgan et al., 
Nat. Phys. 7(1):75-70 (2010)
\bibitem{Budrikis} Z. Budrikis et al., 
Phys. Rev. Lett 109 (30) 037203 (2012)
\bibitem{Branford} W. R. Branford et al., 
Science, 335(6076):1597-1600 (2012) 
\bibitem{Ryzhkin}I.A. Ryzhkin. 
Zhurnal Ehksperimentalnoj i Teoreticheskoj Fiziki, 128(3):559-566 (2005) 
\bibitem{Castelnovo1} C. Castelnovo, et al.,
Ann. Rev. Condens. Matter Phys., 3(1): 35-55 (2012)
\bibitem{Ladak1} S. Ladak et al., 
Nat. Phys., 6:359-363 (2010)
\bibitem{topor} Y. Lao et al., 
Nature Phys. 14, 723-727 (2018)
\bibitem{Chern2} G.-W. Chern, P. Mellado, 
EPL 114 (3): 37004 (2016)
\bibitem{Le} B. L. Le et al., 
Phys. Rev. B, 95:060405 (2017)
\bibitem{Chern3} G.-W. Chern, 
Phys. Rev. App. 8, 6 : 064006 (2017)
\bibitem{Gliga} S. Gliga, et al., 
Phys. Rev. Lett, 110(11):117205 (2013). 
\bibitem{Nisoli4} C. Nisoli, V. Kapaklis, P. Schiffer, 
Nature Phys.13(3):200-203 (2017)
 \bibitem{Gilbert2} I. Gilbert et al., 
 Nat Phys. 10(9):670-675 (2014)
\bibitem{Bhat} V. S. Bhat et al., 
Phys. Rev. Lett. 111(7):077201 (2013)

\bibitem{gartside} J. C. Gartside et al., 
Nature Nano., 13(1):53-58 (2018)
\bibitem{WangYL2} Y.-L. Wang et al., 
Science 352,  6288: 962-966  (2016)
\bibitem{WangYL} Y.-L. Wang,  et al., 
Nature Nano. 13(7): 560  (2018)
\bibitem{Biamonte}  J. D. Whitfield et al 2012 EPL 99 57004
\bibitem{Jurcisinova} E. Jurcisinova, M. Jurcisin, and A. Bobak, Phys. Lett. A 377,2712 (2013)
\bibitem{vavassori0} A. López-Ortega et al., 
Light: Science \& Applications 9(1), 2020

\bibitem{vavassori} Z. Li et al.,
Small 14, 1800868 (2018)

\bibitem{GilbertMem}  I. Gilbert et al., 
Phys. Rev. B, 92(10):104417 (2015) 
\bibitem{Lammert2} P. E. Lammert et al., 
Nat. Phys., 6(10):786-789 (2010)
\bibitem{chialvo} D. Chialvo, 
Nature Phys. 6, 10 744-750 (2010)
\bibitem{diventra} F. L. Traversa, M. Di Ventra, IEEE Tran.  on Neur. Netw. and Learn. Sys. 26(11) pp 2702-2715 (2015)
\bibitem{traversa} F. Traversa, M. Di Ventra,  
J. App. Phys. 123 (2018)

\bibitem{memr1} F. L. Traversa et al., 
Science Advances  1 (6), e1500031 (2015)
\bibitem{bernstein1} A. Imre et al.,
Science, 311 (5758) 205-208,  (2006).
\bibitem{bernstein2} G. Csaba, et al.,
IEEE Trans. on Nano., 99(4), 2009 (2003)
\bibitem{bernstein3} M. Gonellia, et al., 
J. of Magnetism and Magnetic Materials 460, 432 (2018)
\bibitem{Arava} H. Arava, et al. 
Nanotechnology 29, no. 26 265205  (2018)
\bibitem{CASI} J. H. Hensen, E. Folven, G. Tufte, 
Proc. of ALIFE 2018, pp. 15-22, MIT Press, 10.1162/isal-a-00011 (2018)
\bibitem{bernstein4} M. T. Niemier, et al.,
Journal of Physics: Condensed Matter, 23(49), 493202 (2011)
\bibitem{Gypens} P. Gypens, J. Leliaert, and B. Van Waeyenberge ,
Phys. Rev. Applied 9, 034004 (2018)
\bibitem{Ke} X. Ke et al., 
Appl. Phys. Lett. 93, 252504 (2008)
\bibitem{Zhang2} S. Zhang et al.,  
Phys. Rev. Lett 109(8), p.087201 (2012)
\bibitem{ProbComp0} A. A. Faisal, L. P. Selen, D. M. Wolpert,  
Nature reviews. Neuro. 9(4):292 (2008)
\bibitem{redner} P. Krapivsky, S. Redner, S., E. Ben-Naim, A Kinetic View of Statistical Physics. Cambridge: Cambridge University Press. doi:10.1017/CBO9780511780516, (2010)
\bibitem{youtube} \href{https://www.youtube.com/watch?v=ugAiZ0O7tv0}{https://www.youtube.com/watch?v=ugAiZ0O7tv0}
\bibitem{Baxter}  R. Baxter, Exactly Solved Models in Statistical Mechanics, Academic Press (London), 1989
\bibitem{Kibble}  T. W. B.  Kibble,  
J. Phys. A: Math. Gen. 9: 1387 (1976)
\bibitem{Zurek} W. H. Zurek, 
Acta Phys. Pol. B. 24: 1301  (1993)
\bibitem{RFIM} R. Dobrin, J. H. Meinke, P. M. Duxbury, 
J. Phys. A: Math. Gen. 35 L247(2002)
\bibitem{YL} T. D. Lee, C. N. Yang,  
Phys. Review, 87: 410- 419 (1952)





\end{thebibliography}

\appendix
\clearpage
\begin{centering}
\Large
\textbf{Appendix}
\end{centering}
 \section{Integration of  Boolean Gates in Tree-Like Circuits}

\subsection{Gates}

Let us consider the case in which there is no the interaction among the inputs of each gate. We will see later, both numerically and analytically, that, when added, such interaction  does not completely compromise computation, though it can create defects that affect both convergence and reliability. Such gate is also the simplest possible tree and therefore we will introduce here nomenclatures and methods to be used later. The energy of the gate is given as

\begin{equation}
H_3=\sigma_3\left( J(\sigma_1+\sigma_2)+h\right),
\label{eq:paramH}
\end{equation}
where $J$ is the coupling constant between inputs and output and $h$ describes the effect of the biasing moment. Even though we chose the same coupling $J$ for two interactions, all the AND, OR, NAND  and NOR gates can be realized by that choice. Indeed, the functionality of the gate is dictated by the value of the output spin $\sigma_3$, which can be -1 or 1, that minimizes the energy, given fixed values  of the input spins $\sigma_1$, $\sigma_2$. It is thus immediate to show that  gates AND, OR, NAND  and NOR correspond to different quadrants of the $J, h$ plane, given by the condition
\begin{equation}
2 |J|>|h|,
\label{eq:2Jh}
\end{equation}
(where $\sigma=+1$ is TRUE  and $\sigma=-1$ is FALSE). Within that request we have: AND gates (respectively OR)  when $J<0$, $h>0$ ($h<0$ resp.) and NAND gates (NOR) when $J>0$, $h<0$ ($h>0$ resp.). This is represented in Fig. \ref{fig:paramH}.

We also provide evidence of how boolean trees can be integrated in  tree-like circuits. We will use Monte Carlo methods to show that the system converges to a ground state, and will use exact results on the partition function to connect the slowness in convergence for the case with triangular interactions to the emergence of a phase transition.

Our assumption is that the exchange coupling can be both negative and and positive. Ideally, we assume a dipole-dipole interaction (DDI) between magnetic nanoislands but the method we develop applies to arbitrary systems.  Between two magnetic islands $i$, $j$,
provided the vector $\vec r_{ij}$ between the center of the two islands, the absolute distance is given by $r_{ij}=|\vec r_{ij}|$. The energy associated with the interaction can be written in terms of dipole direction and the vector $\vec r_{ij}$, as:
\begin{equation}
H_{ij}=-\frac{\mu}{4\pi r_{ij}^3}\left(\hat \sigma_i \cdot \hat \sigma_{j} -3 (\hat \sigma_i \cdot \hat r_{ij})(\hat \sigma_{j} \cdot \hat r_{ij}) \right),
\label{eq:ferrint}
\end{equation}
which in turn can be written in terms of the angles $\theta_1$ and $\theta_2$ respect to the direction $\hat r_{ij}$ between the centers of the spins:
\begin{equation}
H_{ij}=-\frac{\mu}{4\pi r_{ij}^3}\left(\cos(\theta_i-\theta_j) -3 \cos(\theta_i) \cos(\theta_j)\right) \sigma_i \sigma_{j}.
\end{equation}
In the equation above, we have introduced the scalar spin variables $\sigma_i$ and $\sigma_{j}$.  Let us consider now the case where all the spins are pointing upwards and focus on the case of the interaction between the bottom and the top spin of the triangles. In this case $\theta_1=\theta_2$.  As a convention, the bottom spins are $\sigma_1$ and $\sigma_2$, and the top spin is $\sigma_3$.
It is then easy to see that we can write $\theta_1=\theta_2\equiv\theta$ as a  
function of $r$ and $h$, as $\theta=\arctan(\frac{2r}{l})$, and thus:
\begin{equation}
H_{i3}=-\frac{\mu}{4\pi r^3}\left(1 -3 \sin\left(\arctan\left(\frac{2r}{l}\right)\right)^2 \right) \sigma_i \sigma_3
\label{eq:trianH}
\end{equation}
and as $r\rightarrow \infty$, $\theta\rightarrow0$. We see that there is a specific angular dependence on the coupling sign.


We see that for $r=r^*=\frac{\sqrt{2} l}{3}$ the interaction parameter is zeros, and for $r>r^*$, the interaction changes sign.
Regarding the horizontal spins, we have $\sigma_i \cdot r_{12} =0 $ for $i=1,2$, and the energy
\begin{equation}
H_{ij}=-\frac{\mu}{4\pi r^3}\sigma_i \sigma_{j}.
\label{eq:trianH2}
\end{equation}
{In the bulk of the paper we  exploit the fact that in the case of magnetic nanoislands, which possess a dipole moment,  it is possible engineer the position and direction of the dipoles in order to choose the sign and the interaction strengths.}

\subsection{Gates with horizontal interactions}
\label{sec:3s3int}
{The first generalization we perform is to add horizontal interactions between spin $\sigma_1$ and $\sigma_2$ in the tree model introduced.  As a first generalization, we consider the case in which vertical interactions and horizontal interactions have the same coupling. In the next sections we consider the case in which the spins interact on isosceles triangles on the plane. This means that vertical interactions are identical, but because of the property of the property of dipole interactions described above, the horizontal interaction can have a different value related to the relative angle between the spins.  Such inclusion is also in view of plausible experimental implementations in which horizontal interactions are hard to suppress.
We report for completeness the table of truths for the AND, OR, XOR, NAND and NOR in Tab. \ref{tab:truthv}.}

\begin{table}
\center
\begin{tabular}{|c|c|c|c|c|c|c|}
\hline
$\sigma_1$ & $\sigma_2$ & AND & OR & XOR & NAND & NOR\\ 
\hline
$-1$ & $-1$ & $-1$   & $-1$ & $-1$ & $1$ & $1$\\
\hline 
$-1$ & $1$ & $-1$ & $1$ & $1$  & $1$ & $-1$\\
\hline 
$1$ & $-1 $ & $-1$ & $1$ & $1$  & $1$ & $-1$\\
\hline 
$1$ & $1$ & $1$ & $1$  & $-1$ & $-1$ & $-1$\\
\hline
\end{tabular}
\caption{Table of truth: we assume that spin down corresponds to a false value, meanwhile a spin down to a true value.}
\label{tab:truthv}
\end{table}

\begin{table*}
\center
\begin{tabular}{|c|c|c|c|c|c|c|}
\hline
$\sigma_1$& $\sigma_2$ &$H_{12}$ & $H_{123}(+1)$ & $H_{123}(-1)$ & $H_{12}+H_{123}(+1)$ & $H_{12}+H_{123}(-1)$ \\
\hline
$-1$ & $-1$ &$ J-2h$ & $-2J+h$ & $-h+2J$ & $-h-J$ & $3J-3h$ \\
\hline
$-1$ & $+1$ &$-J$ & $h$ & $-h$ & $h-J$ & $-h-J$ \\
\hline
$1$ & $-1$ &$-J$ &  $h$ & $-h$ & $h-J$ & $-h-J$ \\
\hline
$1$ & $1$  &$J+2h$ & $h+2J$ & $-h-2J$ & $3J+3h$ & $h-J$\\
\hline
\end{tabular}
\caption{We observe that the addition of the horizontal interaction implies a non-trivial shift to the energies.}
\label{tab:intshift}
\end{table*}

The Ising model description of the interaction between the two models is given by:
\begin{eqnarray}
H^3(\vec \sigma)&=&J \sum_{i,j} \sigma_i \sigma_{j}+h \sum_i \sigma_i=J(\sigma_1 \sigma_2+\sigma_1 \sigma_3+\sigma_2 \sigma_3) \nonumber \\
&+&h (\sigma_1+\sigma_2+\sigma_3), \nonumber \\
&=& \underbrace{\sigma_1 \sigma_2 J + h(\sigma_1+\sigma_2)}_{H_{12}}+\underbrace{J \sigma_3 (\sigma_1+\sigma_2)+ h \sigma_3}_{H_{123}} \nonumber\\
&=& H_{12}+H_{123},
\label{eqn:h3}
\end{eqnarray}
which are presented in Tab. \ref{tab:intshift} as a function of the spins $\sigma_1$ and $\sigma_2$.
We are interested in the behavior of the spin $\sigma_3$, which we interpret as the output of the interaction, given $\sigma_1$ and $\sigma_2$. We write $E(\sigma_3| \sigma_1,\sigma_2)$ as the energy of the Hamiltonian of  eqn. (\ref{eqn:h3}) given the values of $\sigma_1,\sigma_2$, and as a function of $\sigma_3$. We can thus construct the Table \ref{tab:treemodel}, which shows how the energy gap emerges.

\begin{table}
\center
\begin{tabular}{|c|c|c|c|}
\hline
$\sigma_1$ & $\sigma_2$ & $E(\sigma_3=1|\sigma_1,\sigma_2)$ & $E(\sigma_3=-1|\sigma_1,\sigma_2)$ \\ 
\hline
$-1$ & $-1$ & $-h -J$ & $3J-3h$ \\
\hline 
$-1$ & $1$ & $h-J$ & $-h-J$ \\
\hline 
$1$ & $-1 $ & $h-J$ & $-h-J$ \\
\hline 
$1$ & $1$ & $3h+3J$ & $h-J$ \\
\hline
\end{tabular}
\caption{Energies for the 3-spins with horizontal interaction}
\label{tab:treemodel}
\end{table}

Let us assume that we interpret the interaction above as a gate where $\sigma_1,\sigma_2$ are the inputs, and $\sigma_3$ is the result, where $-1$ is interpreted as FALSE and $+1$ as TRUE. We ask whether which type of gates can be encoded in the ground state. 


We ask whether again, given the parameters $J$ and $h$, it is possible to obtain such table of truth in the ground state.
For instance, for the gate AND, we ask whether there is region in the plane $(J,h)$ in which:
\begin{eqnarray}
AND: -h-J&\geq& 3J-3h \nonumber \\
h-J&\geq& -h-J  \nonumber  \\
3h+3J &\leq & h-J
\label{eq:ANDen}
\end{eqnarray}
and for the gate OR:

\begin{eqnarray}
OR: -h-J&\geq& 3J-3h \nonumber \\
h-J&\leq& -h-J  \nonumber  \\
3h+3J &\leq & h-J
\label{eq:ORen}
\end{eqnarray}
and finally, the gate XOR:
\begin{eqnarray}
XOR: -h-J&\geq& 3J-3h \nonumber \\
h-J&\leq& -h-J  \nonumber  \\
3h+3J &\geq & h-J.
\label{eq:XORen}
\end{eqnarray}
{The result is the one discussed in the bulk of the paper, and presented in Fig. \ref{fig:paramH}, while the energies are presented in Tab. \ref{tab:3benergy}. We observe that meanwhile the gate XOR  conditions are not all feasible, the gates AND and OR are feasible. Both of them require $J<0$, i.e. ferromagnetic interaction, and for $h>0$ we obtain an AND gate, meanwhile for $h<0$ we obtain an OR gate. The situation is instead inverted for $J>0$. This suggests that the local field $h$ can program the system to obtain a behave as OR or AND gate. This picture suggests the use of external field to reprogram, given the sign of $J$, the type of gate one aims to use. }

{The approach we use works if we pin the input spins $\sigma_1$ and $\sigma_2$. The key problem when all $\sigma$'s are free to fluctuate with this approach in general, is that there is strong degeneracy, and the minimum state is given by $-1,-1,-1$ for $J\leq 0$.
We can necessary to try to split the degeneracy by considering two set of couplings as mentioned before $J_{12}$ for the coupling between spin $\sigma_1$ and $\sigma_2$ and $J_3$ between spins $\sigma_1$ and $\sigma_2$ with $\sigma_3$. 
We thus consider the following Hamiltonian:
\begin{equation}
H=J_{12} \sigma_1 \sigma_2 +J_3 \sigma_3(\sigma_1+\sigma_2)+h_{12} (\sigma_1+\sigma_2)+h_3 \sigma_3
\end{equation}
and this Hamiltonian better represents the dipole interaction between not uniaxial dipoles. The energy states are presented in Tab. \ref{tab:3benergy} forthe $8$ states between the three spins.}
 \begin{table}
 \center
\begin{tabular}{|c|c|c|c|}
\hline
$\sigma_1$ & $\sigma_2$ & $\sigma_3$ & $E$ \\ 
\hline
$-1$ & $-1$ & $-1$ & $J_{12}+2 J_3-2 h_{12}-h_3$  \\
\hline 
$-1$ & $-1$ & $1$ &  $J_{12}-2 J_3-2 h_{12}+h_3$ \\
\hline 
$-1$ & $1 $ & $-1$ &  $-J_{12}-h_3$\\
\hline 
$-1$ & $1$ & $1$ &  $-J_{12}+h_3$\\
\hline
$1$ & $-1$ & $-1$ &  $-J_{12}-h_3$\\
\hline 
$1$ & $-1$ & $1$ &  $-J_{12}+h_3$\\
\hline 
$1$ & $1 $ & $-1$ &  $J_{12}-2J_3+2h_{12}-h_3$\\
\hline 
$1$ & $1$ & $1$ &  $J_{12}+2 J_3 +2 h_{12}+h_3$\\
\hline
\end{tabular}
\caption{Energy values for the various states of a 3-spins Hamiltonian.}
\label{tab:3benergy}
\end{table}

{In the previous calculation for the average of $\sigma_3$, the output spin, we have used the couplings $J$'s between the spins to calculate the average. The same can be done, however, by the mentioned star-triangle transformation between the spins, thus the couplings $n$'s. As mentioned, this mapping introduces an unphysical spin $\sigma_{\blacke}$ which directly couples to the three physical spins, which we call now $\sigma_{\blackf}$.}

The effective Hamiltonian for the interactions is given by:
\begin{equation}
H/T=L_{12} (T)\sigma_{\blacke}(\sigma_{\blackf}^1+\sigma_{\blackf}^2)+L_{3}(T) \sigma_{\blacke} \sigma_{\blackf}^3+ h(\sigma_{\blackf}^1+\sigma_{\blackf}^2+\sigma_{\blackf}^3)
\end{equation}
where $h$ is normalized in the temperature. Let us consider now the average $\langle \sigma^3_{\blackf}\rangle$ as a function of the temperature for these couplings. The  average output spin is given by:
\begin{eqnarray}
\langle \sigma^3_{\blackf}\rangle=\frac{Z(\sigma_{\blackf}^3=+1)-Z(\sigma_{\blackf}^3=1)}{Z(\sigma_{\blackf}^3=+1)+Z(\sigma_{\blackf}^3=1)}.
\end{eqnarray}

\section{Partition function recursion}
\label{sec:partfunc}
{We are not interested in deriving a recursion relation for the partition function of the system both for $J_{12}=0$ and $J_{12}\neq 0$, via the star-triangle transformation. The importance of such recursion relation is that it simplifies the numerical study of the Lee-Yang zeros. We use the formalism introduced for the star-triangle transformation of the physical $\sigma_{\blackf}$ and the unphysical $\sigma_{\blacke}$ spins.}
We define $Z(\sigma^r_{\blackf},h^\prime)=\sum_{\sigma^r_{\blackf}} e^{h^\prime \sigma^r_{\blackf} } \tilde Z(\sigma^r_{\blackf}) $.
We also consider for simplicity $h$ homogeneous for the time being. If the root is the spin $\sigma_{\blackf}^0$, then we have:

\begin{eqnarray}
Z(\sigma^0_{\blackf},\tilde h)&=&\sum_{\sigma^l_{\blacke}} \sum_{\sigma^0_{\blackf}} e^{ \sigma^0_{\blackf}(\tilde h+\tilde L_3 \sigma^l_{\blacke})} \cdot \sum_{\sigma^1_{\blackf},\sigma^2_{\blackf} }e^{\tilde L_{12} \sigma^l_{\blacke} (\sigma^1_{\blackf}+\sigma^2_{\blackf})+\tilde h(\sigma^1_{\blackf}+\sigma^2_{\blackf})} \tilde Z(\sigma^1_{\blackf}) \cdot \tilde Z(\sigma^2_{\blackf})\nonumber \\
&=&\sum_{\sigma^l_{\blacke}} 2 \cosh(\tilde h+\tilde L_3 \sigma^l_{\blacke})\sum_{\sigma^1_{\blackf},\sigma^2_{\blackf} } e^{\tilde L_{12} \sigma^l_{\blacke} (\sigma^1_{\blackf}+\sigma^2_{\blackf})+\tilde h(\sigma^1_{\blackf}+\sigma^2_{\blackf})} \tilde Z(\sigma^1_{\blackf}) \cdot \tilde Z(\sigma^2_{\blackf}) \nonumber \\
&=&2 \sum_{\sigma^1_{\blackf},\sigma^2_{\blackf} }\left( \cosh(\tilde h+\tilde L_3) e^{\tilde L_{12}  (\sigma^1_{\blackf}+\sigma^2_{\blackf})}+\cosh(\tilde h-\tilde L_3) e^{-\tilde L_{12}  (\sigma^1_{\blackf}+\sigma^2_{\blackf})}  \right) \nonumber \\
& & \ \ \ \ \ \ \ \ \ \ \ \ \ \ \cdot e^{\tilde h(\sigma^1_{\blackf}+\sigma^2_{\blackf})} \tilde Z(\sigma^1_{\blackf}) \cdot \tilde Z(\sigma^2_{\blackf}) \nonumber \\
&=&2\sum_{\sigma^1_{\blackf},\sigma^2_{\blackf} }   \cosh(\tilde h+\tilde L_3) e^{\tilde L_{12}  (\sigma^1_{\blackf}+\sigma^2_{\blackf})} e^{\tilde h(\sigma^1_{\blackf}+\sigma^2_{\blackf})} \tilde Z(\sigma^1_{\blackf}) \cdot \tilde Z(\sigma^2_{\blackf}) \nonumber \\
&+& 2 \sum_{\sigma^1_{\blackf},\sigma^2_{\blackf} }\cosh(\tilde h-\tilde L_3) e^{-\tilde L_{12}  (\sigma^1_{\blackf}+\sigma^2_{\blackf})}  e^{\tilde h(\sigma^1_{\blackf}+\sigma^2_{\blackf})} \tilde Z(\sigma^1_{\blackf}) \cdot \tilde Z(\sigma^2_{\blackf}) \nonumber \\
&=& 2 \cosh(\tilde h+\tilde L_3)  \sum_{\sigma^1_{\blackf} }  e^{(\tilde L_{12}+\tilde h) \sigma^1_{\blackf}} \tilde Z(\sigma^1_{\blackf}) \sum_{\sigma^2_{\blackf} } e^{(\tilde L_{12}+\tilde h)\sigma^2_{\blackf}} \tilde Z(\sigma^2_{\blackf}) \nonumber \\
&+& 2 \cosh(\tilde h-\tilde L_3)  \sum_{\sigma^1_{\blackf} }  e^{(-\tilde L_{12}+\tilde h) \sigma^1_{\blackf}} \tilde Z(\sigma^1_{\blackf}) \sum_{\sigma^2_{\blackf} } e^{(-\tilde L_{12}+\tilde h)\sigma^2_{\blackf}} \tilde Z(\sigma^2_{\blackf}) \nonumber \\
&=& 2 \cosh(\tilde h+\tilde L_3)  Z(\sigma^1_{\blackf},\tilde h+\tilde L_{12}) Z(\sigma^2_{\blackf},\tilde h+\tilde L_{12}) \nonumber \\
&+&2 \cosh(\tilde h-\tilde L_3)  Z(\sigma^1_{\blackf},\tilde h-\tilde L_{12}) Z(\sigma^2_{\blackf},\tilde h-\tilde L_{12})\nonumber
\end{eqnarray}

It is easy to see that such recursion relation can be generalized. It is simply necessary to replace $h^\prime$ with $h^\prime_{\sigma_{\blackf^j}}$. Thus we have:

\begin{eqnarray}
Z^{n}(\sigma^0_{\blackf},{\tilde h}_{\sigma^0_{\blackf}})&=& 2 \cosh({\tilde h}_{\sigma^0_{\blackf}}+\tilde L_3) \nonumber \\
&\cdot& Z^{n-1}(\sigma^1_{\blackf},{\tilde h}_{\sigma^1_{\blackf}}+\tilde L_{12}) Z^{n-1}(\sigma^2_{\blackf},{\tilde h}_{\sigma^2_{\blackf}}+\tilde L_{12}) \nonumber \\
&+&2 \cosh({\tilde h}_{\sigma^0_{\blackf}}-\tilde L_3)  \nonumber \\
&\cdot&Z^{n-1}(\sigma^1_{\blackf},{\tilde h}_{\sigma^1_{\blackf}}-\tilde L_{12}) Z^{n-1}(\sigma^2_{\blackf},{\tilde h}_{\sigma^2_{\blackf}}-\tilde L_{12}) \nonumber \\
\label{eq:recrel}
\end{eqnarray}

\begin{figure}[t!]
\center
\includegraphics[scale=2.5]{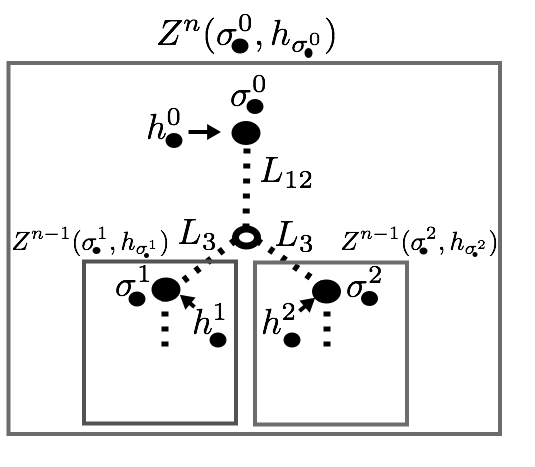}
\caption{The recursive structure of eqn. (\ref{eq:recrel}): $Z^n(\sigma,h)$ represents the partition function rooted at the spin $\sigma$ with external field $h$, which depends on the partition functions at the lowest order $Z^{n-1}$ rooted at the two sub-branches spins.}
\label{fig:refroot}
\end{figure}

In the homogeneous case, this equation can be written explicitly. If $h$ is homogeneous, then the recursion can be written as
\begin{eqnarray}
Z^{n}(\sigma^0_{\blackf},h)&=& 2 \cosh({\tilde h}_{\sigma^0_{\blackf}}+\tilde L_3)  \left(Z^{n-1}(\sigma^1_{\blackf},h) \right)^2\nonumber \\
&+&2 \cosh({\tilde h}_{\sigma^0_{\blackf}}-\tilde L_3)  \left(Z^{n-1}(\sigma^1_{\blackf},h) \right)^2
\label{eq:cumtree}
\end{eqnarray}
It is easy to see that the partition function is invariant under the transformation $\tilde L_3\rightarrow -\tilde L_3,\tilde L_{12}\rightarrow -\tilde L_{12}$. This implies that the zeros of the purely ferromagnetic and antiferromagnetic partition functions are identical.
Also, eqn. (\ref{eq:cumtree}) contains only physical spins, and thus the symbol $\sigma^i_{\blackf}$ can be replaced with $\sigma$ in the formula above, as it appears in the paper. Graphically, the recursion relation is presented in Fig. (\ref{fig:refroot}).

The recursion ends with $Z^1(x)=2 \cosh(x)$. We note that if we define $V_{\blackf}(n)=2^{2^{n-1}-1}$, $V_{\blacke}(n)=2^{2^n-1}$ and the recursive function $F(n)=2 F(n-1)^2, F(1)=1$, then we can study the recursion relation for the rescaled partition function $\mathcal Z^{n}(h)=\frac{F(n)}{V_{\blackf}(n) V_{\blacke}(n)} Z^{n}(h)$, which satisfies
\begin{eqnarray}
 Z^{n}(h)&=&\cosh(h-\tilde L_3) Z^{n-1}(h-\tilde L_{12})^2 \nonumber \\
&+&\cosh(h+\tilde L_3)\mathcal Z^{n-1}(h+\tilde L_{12})^2 \nonumber \\
 Z^{1}(x)&=&\cosh(x),
\end{eqnarray}
and which trivially generalizes to the nonhomogeneous case.
\subsection{Case $L_{12}=0$}
If we assume that ${\tilde L}_{12}=0$, then it further simplifies to:
\begin{eqnarray}
 Z^{n}(\sigma^0_{\blackf},h)&=&  \left(\cosh({\tilde h}_{\sigma^0_{\blackf}}+\tilde L_3) +\cosh({\tilde h}_{\sigma^0_{\blackf}}-\tilde L_3)\right) \nonumber \\
&\cdot&\mathcal Z^{n-1}(\sigma^1_{\blackf},h)^2.
\label{eq:cumtree2}
\end{eqnarray}
If we introduce the variable $y=e^h$, then it is easy to see that $\cosh({\tilde h}_{\sigma^0_{\blackf}}+\tilde L_3)$ can be written
as
\begin{eqnarray}
 Z^{n}(\sigma^0_{\blackf},y)&=&   \frac{(1+y^2) \cosh(\tilde L_3)}{ y}  \mathcal Z^{n-1}(\sigma^1_{\blackf},y) ^2.
\label{eq:cumtree2}
\end{eqnarray}
Let us assume that the zeros of $\mathcal Z^{n-1}$ are $y_1^{n-1},\cdots,y_k^{n-1}$, and thus $\mathcal Z^{n-1}=a_{n-1} \prod_{i=1}^k (y-y^{n-1}_i)$.
We have:
\begin{eqnarray}
 Z^{n}(\sigma^0_{\blackf},y)&=& 2 \frac{(1+y^2) \cosh(\tilde L_3)}{ y}  a_{n-1}^2 \prod_{i=1}^k (y-y^{n-1}_i)^2.\nonumber
\end{eqnarray}
and thus we have that the zeros at the order $n$ will have two further zeros:
\begin{equation}
y^\pm=\pm i 
\end{equation}
which are imaginary.  
We note that this is independent from the sign of $\tilde L_3$. In the case the external field is not homogeneous the result easily generalizes. In fact, we have:
\begin{eqnarray}
Z^{n-1}(\sigma^1_{\blackf},{\tilde h}_{\sigma^1_{\blackf}})&=&a_{n-1}^1 \prod_{i=1}^k (y-a^{n-1}_i) \nonumber \\
Z^{n-1}(\sigma^2_{\blackf},{\tilde h}_{\sigma^2_{\blackf}})&=&a_{n-1}^2 \prod_{i=1}^k (y-b^{n-1}_i)  \nonumber
\end{eqnarray}
and thus if $Z^{n-1}$ does not have any real zeros neither will $Z^n$. The result follows from noticing that $Z^0(y)=\frac{1+y^2}{2y}$, which does not have any real zeros.
We thus see that the model with $L_{12}=0$ cannot have any phase transition. 

\subsection{Case $\tilde L_{12}\neq0, L_3>0$}
Let us now consider the more interesting case in which $L_{12}\neq0$. In this case, the model is not exactly a tree anymore.

\begin{figure}
\center
\includegraphics[scale=0.4]{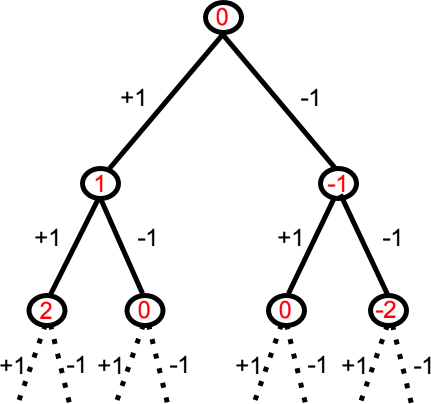}
\caption{The combinatorial structure emerging from eqn. (\ref{eq:cumtree}).}
\label{fig:cumtree}
\end{figure}

From the combinatorial point of view, the terms which appear in the partition function can be obtained from the analysis of the tree in Fig. \ref{fig:cumtree}. Each branch of the tree is the product of factors 
\begin{eqnarray}
q(x)&=&2\cosh(h+f L_3 +\Sigma(x) L_{12})\nonumber \\
&=& 2\Big(\cosh(h)\cosh(f L_3 +\Sigma(x) L_{12}) \nonumber \\
&+& \sinh(h)\sinh(f L_3 +\Sigma(x) L_{12})\Big) \nonumber \\
&=& \frac{e^{-f L_3-L_{12} \Sigma(x) }+y^2 e^{f L_3+L_{12} \Sigma(x) }}{2 y}
\end{eqnarray}
with $f=\pm 1,0$, and where $\Sigma(x)$ is the number in the circle for each branch. These factors can be obtained summing over each possible branch of the tree.
Let us consider the possible combinations of strings of length $n$ made of $\pm1$, $\sigma^n=(\sigma_1,\sigma_2, \cdots,\sigma_n)$. Given a combination $\mathcal B=\{\sigma(1),\cdots,\sigma(n)\}$, we construct the following sequence of scalar based on the cumulative sum, $\Sigma^{\mathcal B}(1)=0$,$\Sigma^{\mathcal B}(n+1)=\sum_{i=1}^n \sigma^{\mathcal B}(i)$. 
This implies that the partition function is a multinomial in the fundamental variables

We thus have the following set of strings:
\begin{equation}
C^{\mathcal B}=\left(  \begin{array}{cccc}
 \sigma_1& \sigma_2 & \cdots &\sigma_n \\
\Sigma^{\mathcal B}(1) &\Sigma^{\mathcal B}(2) & \cdots&\Sigma^{\mathcal B}(n)
\end{array} \right)
\end{equation}
indexed by the branch ${\mathcal B}$. The recursion relation cannot be solved exactly, and thus we solved it numerically.

\end{document}